\documentclass[cameraready]{Interspeech}
\usepackage{comment}

\title{Unsupervised Approaches for Global Prosodic Embedding Extraction}

\author[affiliation={1}, orcid=0009-0003-3042-9385]{Martin}{Meza}
\author[affiliation={2}, orcid=0000-0002-0426-8683]{Luciana}{Ferrer}
\author[affiliation={2}, orcid=0000-0001-8570-4688]{Pablo}{Riera}

\address{
    $^1$ Departamento de Computación, FCEyN, Universidad de Buenos Aires (UBA), Argentina \\
    $^2$ Instituto de Investigación en Ciencias de la Computación (ICC), CONICET-UBA, Argentina 
}

\email{mmeza@dc.uba.ar, lferrer@dc.uba.ar, priera@dc.uba.ar}


\begin{document}

\maketitle

\begin{abstract}
Prosody is central to oral communication, conveying information like the emotional state of the speaker and cues needed for meaning disambiguation. Many self-supervised models of speech produce embeddings that encode prosodic as well as linguistic, and speaker information. This entanglement of information is problematic in scenarios where prosody is the main distinguishing factor while other factors may vary between training and deployment; in such cases, a purely prosodic representation would be more robust. Such representation could also be used for analyzing the role of prosody in a given task or as input to speech synthesis systems. In this work, we propose a variety of approaches for producing global prosodic embeddings based on auto-encoder models of pitch and energy. We develop a benchmark for assessing the performance of these representations, showing that our embeddings provide competitive or superior performance under challenging conditions, compared to various alternatives.
\end{abstract}

\section{Introduction}
Prosody plays a central role in spoken communication. Intonation contours distinguish questions from statements, convey emotions, signal discourse structure, and resolve pragmatic ambiguities~\cite{Ladd2008}. Most modern speech processing systems represent prosody implicitly. Large self-supervised models such as wav2vec 2.0~\cite{wav2vec}, HuBERT~\cite{hubert}, and WavLM~\cite{wavlm} learn powerful speech representations which encode prosodic information along with speaker identity, linguistic content, and recording conditions \cite{prosaudit, superb-prosody}. 
The entanglement of prosody with other aspects of the speech signal makes these embeddings suboptimal, fragile, or unnecessarily complex for tasks like disambiguation, in which the only distinguishing aspect is prosody \cite{desambiguation, desprosody}. Even when other information may be useful to solve the task, like in emotion classification where both linguistic content and prosody complement each other, the linguistic content may tend to vary between development and deployment data, causing a model that encodes linguistic information to degrade in practice. 
Embeddings that are purely prosodic could be used as robust representations for such applications.

A purely prosodic representation may also be used to draw conclusions about the importance of prosody for a given task. 
Ablation studies adding prosodic embeddings to text- or speaker-based embeddings could shed light into the contribution of each aspect. This has been explored, for example, in the context of neurological disorder detection \cite{alz1, autism, autism2}.
Finally, embeddings that are purely prosodic may improve performance of speech synthesis systems where the text and speaker information are provided separately.


Besides disentanglement, another important requirement for prosodic embeddings is contextualization. High-level prosodic phenomena are thought to be structured over intonation units (IUs) \cite{IU}, discourse-level prosodic units characterized by a coherent melodic contour. Ideally, prosodic representations should be extracted at this level of granularity to properly reflect a complete prosodic pattern. 

Targeted approaches have been proposed to produce IU- or utterance-level representations that capture only prosodic information. Traditional methods for this purpose rely on hand-crafted features given by pitch, energy and speech timing statistics \cite{distrust, hyperarticulate}, which have the advantage of being designed by experts based on linguistic knowledge, and the disadvantage of being potentially suboptimal or incomplete. 
More recent approaches train embedding extraction models, much like the self-supervised models mentioned above, but using pitch and energy signals as input rather than the full waveform, mitigating the chance of entanglement with lexical and speaker cues. 

Among this family of methods, Portes and Hor\'ak~\cite{prosodyVQVAE2024,prosodyVQVAE2025} trains  VQ-VAE models to jointly embed frame-level F0 and energy into a discrete codebook, optimizing for reconstruction error. Their model is evaluated only in terms of reconstruction metrics rather than downstream tasks, leaving open the question of whether the learned codes efficiently capture high-level prosodic information.

Other works that develop  prosodic representations can be found in the context of speech synthesis. Skerry-Ryan et al.~\cite{SkerryRyan} introduced a prosody encoder jointly trained with Tacotron 2 to enable prosody transfer, learning a fixed-dimensional embedding from mel-spectrograms or pitch and energy only. They assume the embeddings capture prosody since speaker identity and phonetic content are already modeled by other components, leaving prosody as the remaining source of variation.  
A subsequent work modeled pitch alone to encode prosodic style for speech synthesis \cite{Lilja2024IsPC}. Another approach proposed by Hodari et al.~\cite{ZackHodari} proposed to learn utterance-level embeddings from pitch contours within a synthesis framework. The resulting embeddings were clustered to reveal interpretable prosodic categories.
More recent synthesis systems model prosody through latent diffusion processes~\cite{NaturalSpeech2}, neural codec language models~\cite{valle}, or explicit style modules~\cite{styletts, megatts}.


In general, prosodic representations designed for speech synthesis are learned end-to-end for the task and are not meant to be general-purpose. Further, these methods do not guarantee that the representations are purely prosodic since, even if text and speaker identity information is introduced to the model, the learned embedding may still encode residual information from those aspects.   
Finally, evaluation on downstream classification tasks is not performed, so it remains unknown whether these models generalize beyond the synthesis pipeline for which they were trained. 

Another method aimed at producing embeddings that contain prosodic information is emotion2vec \cite{emo2vec}. In this approach, a model that takes the full waveform as input is pre-trained for reconstruction on emotional data and further fine-tuned for emotion classification. Since emotion classification is a task that benefits from both prosodic and semantic information, this model is not expected to preserve only prosodic information. 



In this paper, we aim to develop a general-purpose prosody embedding extractor that (i) produces global fixed-length compact representations; (ii) captures prosodic information exclusively, as a consequence being relatively robust to speaker and linguistic changes; and (iii) is learned in a self-supervised manner from prosodic signals alone, without relying on text, speaker labels, or downstream task supervision.
Such an embedding could serve as a drop-in prosodic feature for diverse downstream tasks, from emotion recognition to intonation classification, and could complement text or speaker-based representations in multi-modal systems. 

To this end, we propose and compare various autoencoder architectures, where the prosodic embedding is taken from the output of the encoder. The models take the frame-level F0 and energy signals as input and are trained with a self-supervised approach.  
The main contributions of this work are:
\begin{enumerate}
    \item A novel downstream evaluation methodology. 
    Rather than evaluating on tasks where prosody is confounded with other factors, we focus on tasks for which the main distinguishing aspect between the classes is prosody. To this end, we use datasets where a set of speakers produce the same phrases under different prosody-related conditions, and define three evaluation protocols of increasing difficulty: speaker-independent, speaker-and-text-independent, and a spurious correlation protocol that explicitly tests robustness to linguistic content. This allows us to measure the degree to which a representation preserves prosodic information and its robustness to other factors. 
    \item A new dataset specifically of synthetic speech designed for controlled evaluation of prosodic representations, where all combinations of speaker, text, and prosodic style are available by design. 
    This dataset is used for downstream evaluation, along with two publicly available datasets, using the protocols described above. 
    \item A systematic and extensive comparison of multiple autoencoder architectures for learning fixed-dimensional prosody embeddings from F0 and energy signals alone, exploring the impact of architecture, information bottleneck, and training objective across different input--output configurations. 
    \item A systematic comparison with state-of-the-art approaches including hand-crafted features, large self-supervised speech models, emotion2vec, and a prosodic VQ-VAE baseline \cite{prosodyVQVAE2025}.
\end{enumerate}
We publicly release the new benchmark, trained models and code to foster reproducibility and further research.\footnote{\url{https://github.com/martinBmeza/prosodic_embeddings}}

\section{Method}
We propose a self-supervised framework for learning fixed-dimensional prosody embeddings from frame-level prosodic features: F0, energy, and voicing signals. By restricting the input to be prosodic signals, we greatly limit the amount of speaker and text information that may be encoded in the representations. These input signals are fed into an autoencoder which is trained on unlabeled speech. Fixed-size global embeddings $\mathbf{z} \in \mathbb{R}^d$ are then extracted from the encoder's output. 

All prosodic embedding extractors follow the encoder-decoder paradigm: an input sequence $\mathbf{X} = (\mathbf{x}_1, \dots, \mathbf{x}_T)$, where each $\mathbf{x}_t \in \mathbb{R}^C$ is a frame-level prosodic feature vector, is processed by an encoder to produce a latent representation. Depending on the architecture, this representation may take the form of a single fixed-dimensional vector $\mathbf{z} \in \mathbb{R}^d$,
or a variable-length sequence $\mathbf{Z} = (\mathbf{z}_1, \dots, \mathbf{z}_{T'})$, with $\mathbf{z_i} \in \mathbb{R}^{d'}$. The decoder then reconstructs the input sequence from this latent representation. The model is trained end-to-end by minimizing a reconstruction loss of pitch, energy, and, in some configurations, a voicing signal. Then, only the encoder is used at inference time; the decoder is only used to provide a training signal to encourage the embedding to capture the relevant information for reconstructing the input signals. For downstream tasks, sequential encoder outputs are aggregated into a single fixed-dimensional vector $\mathbf{z} \in \mathbb{R}^d$, ensuring a unified representation regardless of the encoder’s original output structure.

We explore two main families of encoder architectures: recurrent and Transformer-based, studying the impact of several design factors including decoder type, reconstruction target, and pretext task. The following sections describe the architectures and each aspect of our proposed approach in detail.

\subsection{Input Features}
\vspace{-1mm}
The input to all models consists of prosodic features extracted at the frame level. These features are extracted from 16kHz audio waveforms. Waveforms originally sampled at other sampling rates are resampled to 16 kHz before feature extraction. Below, we describe how each feature is computed.

\textbf{Fundamental frequency (F0).} Fundamental frequency is extracted using the Praat \cite{praat, pypraat} autocorrelation method with a hop size of 10 ms; the analysis window length is determined by the minimum F0 frequency setting. The minimum and maximum F0 values are determined using the two-pass method described in \cite{hirst2011analysis}, where an initial broad-range estimation is first obtained and subsequently refined based on the speaker-specific F0 distribution. Unvoiced frames are flagged by the algorithm with a value of zero. A binary voicing sequence is derived from this raw F0 signal before any further processing. The F0 values of voiced frames are then converted to a logarithmic scale, and subsequently linearly interpolated across unvoiced regions to produce a continuous log-F0 signal.

\textbf{Energy.} A perceptually motivated loudness measure is computed following the procedure used in egemaps \cite{egemaps} to extract the loudness low-level descriptor. The signal is segmented into 20 ms Hamming-windowed frames with a 10 ms hop size and converted to the power spectrum via FFT. The power spectrum is projected onto a 26-band mel filterbank (20-8000 Hz), and each band is weighted by an equal-loudness curve evaluated at the band's center frequency. Cubic-root compression is applied to the weighted mel-band energies, and these values are summed across bands to yield a single loudness value per frame. 

Pitch and energy are z-normalized using global statistics computed over the training set. For pitch, the mean and standard deviation are computed over voiced frames only.

\subsection{Auto-encoder architectures}
\vspace{-1mm}

We explore two families of architectures based on recursive neural networks and transformers. In both cases, the input signals are given by the interpolated pitch, energy and voicing frame-level signals. The output signals are pitch and energy. A voicing signal is only produced in some configurations, as explained in the Section~\ref{sec:pretext_task}

\subsubsection{GRU-Based Autoencoder}
The recurrent models use a bidirectional Gated Recurrent Unit (GRU) encoder. The input sequence is first projected to dimension $d$ through a feedforward layer with ReLU activation and dropout, and then processed by a multi-layer bidirectional GRU where all layers have dimension $d$. The embedding $\mathbf{z} \in \mathbb{R}^d$ is obtained by concatenating the final hidden states of the forward and backward directions of the last GRU layer, resulting in a $2d$-dimensional vector, which is then projected back to $\mathbb{R}^d$ through a linear layer. This produces a single fixed-size vector that summarizes the entire input sequence.

The decoder head for pitch and energy is a unidirectional GRU. At each time step $t$, the decoder receives the concatenation of the previous output (or ground-truth frame during teacher forcing, as explained in Section \ref{sec:upstream_training}) and the embedding $\mathbf{z}$, encouraging $\mathbf{z}$ to retain sufficient information to guide reconstruction throughout the sequence. The initial hidden state of the decoder is derived from $\mathbf{z}$ through a learned linear layer.

Two options were explored to produce a voicing output signal: an autoregressive GRU decoder like the one used for the pitch and energy signals but with its own separate parameters, and a non-autoregressive decoder. The non-autoregressive decoder repeats $\mathbf{z}$ across all time steps, adds learned positional embeddings, processes the resulting sequence with three GRU layers, and produces the voicing prediction for each frame in parallel. 
Empirically, the autoregressive and the non-autoregressive formulations yielded comparable performance in terms of reconstruction accuracy as well as downstream task evaluations. 
Consequently, for the models compared in this work, we adopt the non-autoregressive variant due to its faster training time and improved computational efficiency.

\subsubsection{Transformer-Based Autoencoder}
For the Transformer-based models, the input sequence is projected to dimension $d$, as for the recurrent architectures, and sinusoidal positional encodings are added to the resulting sequence. The sequence is then processed by a stack of three Transformer layers, each with eight attention heads and a position-wise feedforward network of dimension $4d$. Dropout rate of 0.1 is applied throughout the Transformer blocks. This encoder produces a contextualized representation at each time step. The decoder is a non-autoregressive Transformer decoder that uses learned positional queries and cross-attention to reconstruct all frames in parallel. It mirrors the encoder configuration, comprising three Transformer layers with eight attention heads and a feedforward dimension of $4d$ and the same dropout rate. Finally, pitch, energy and, in some configurations, voicing signals are produced by a linear layer. 

We explore two variants that differ in how information flows from encoder to decoder and how the downstream embedding $\mathbf{z}$ is obtained.

\textbf{Full-sequence variant}: In this configuration, the decoder attends to the full encoder output sequence via cross-attention, so there is no information bottleneck between encoder and decoder. The fixed-dimensional embedding $\mathbf{z}$ for downstream modeling is obtained by computing the mean and standard deviation over the encoder outputs and concatenating them, yielding a vector of twice the encoder dimension. Since the decoder has access to the entire encoded sequence, reconstruction is relatively easy. Note that, in this variant, the quality of the pooled $\mathbf{z}$ embedding is not directly controlled by the training process. 

\textbf{CLS-tokens variant}: This configuration introduces an explicit information bottleneck by extending the input sequence after the linear projection with a learnable CLS token. After encoding, only the output representation corresponding to the CLS position is retained. This value is repeated for the length of the sequence to be predicted and fed into the decoder, thereby forcing all information required for reconstruction to be compressed into this vector. The CLS-token variant combines the Transformer's ability to model long-range dependencies with a compression constraint that directly encourages informative global embeddings.

\subsection{Pretext Tasks}
\label{sec:pretext_task}
All models are trained to reconstruct prosodic signals from the encoder's output. The decoder can be asked to reconstruct different subsets of prosodic signals, by using a combination of two or three of the following  losses:

\begin{itemize}
    \item $\mathcal{L}_E$: Mean squared error between the predicted and target energy signal, computed over all frames in the sequence.
    \item $\mathcal{L}_{P_v}$: Mean squared error between the  predicted and the target log-F0 signal, computed only over voiced frames as determined by the voicing output sequence. 
    \item $\mathcal{L}_{P_i}$: Mean squared error between the target and the predicted interpolated log-F0, computed over all frames. 
    \item $\mathcal{L}_V$: Binary cross-entropy loss between the predicted and target voicing sequence, computed over all frames.
\end{itemize}

We define three reconstruction target configurations, each selecting a different combination of these losses and their associated output signals:

\begin{itemize}
    \item \textbf{EPv}: $\mathcal{L} = \mathcal{L}_{P_v} + \mathcal{L}_E$. The decoder reconstructs pitch and energy. The pitch loss is restricted to voiced frames, so the decoder is not required to predict pitch over unvoiced regions. As a consequence, it is also not required to learn where the unvoiced regions are located.
    \item \textbf{EPi}: $\mathcal{L} = \mathcal{L}_{P_i} + \mathcal{L}_E$. The decoder reconstructs the interpolated pitch and energy signals. The pitch loss involves all frames, encouraging the model to learn a smooth prosodic representation. As in the previous case, the model is not required to learn voicing information.
    \item \textbf{EPvV}: $\mathcal{L} = \mathcal{L}_{P_v} + \mathcal{L}_E + \mathcal{L}_V$. The decoder reconstructs pitch over voiced frames only, energy, and the voicing sequence. This configuration combines the voiced-only pitch loss with an explicit voicing reconstruction term, requiring the encoder to capture both the melodic contour and the voicing pattern. Voicing patterns may indirectly encode speech rate, which is an important part of prosody. 
\end{itemize}
The loss determines the output signals that the decoder is required to produce: energy and pitch for EPv, energy and interpolated pitch for EPi, and energy, pitch and voicing sequence for EPvV.  In all cases, a pitch signal over all frames is produced by the model, but, for EPv and EPvV, the values over unvoiced frames are ignored when computing the loss. 

In addition, we consider two reconstruction tasks: a standard autoencoder reconstruction and a masked auto-encoder reconstruction. In {\bf standard reconstruction}, the  encoder receives the full input sequence and the decoder reconstructs it entirely. The training signal encourages the encoder to produce representations (whether a single vector or a sequence) that retain sufficient information to recover the original prosodic features. 

In \textbf{masked reconstruction}, inspired by Masked Autoencoders (MAE) \cite{MAE}, a fraction of the input frames is masked before encoding, forcing the model to reconstruct missing segments from context. 
The masking procedure is controlled by two hyperparameters: a masking ratio $p$ (the fraction of frames to mask) and a mask segment size $s$ (the number of consecutive frames in each masked span). At each training step, starting positions are sampled uniformly at random along the temporal axis, and spans of $s$ frames are masked until the total number of masked frames reaches approximately a fraction $p$ of the sequence length. After the input features are projected by the input layer to the model dimension~$d$, the mask is applied to the projected sequence, and the masked sequence is fed into the encoder (keeping the masked features or dropping them, depending on the architecture, as explained below). The decoder must reconstruct the full original sequence, including the masked positions, from the encoder's output. At inference time, no masking is applied to extract the embeddings. 

For the GRU-based architectures, masked frames are replaced by a learnable \texttt{[MASK]} token vector $\mathbf{m} \in \mathbb{R}^d$, so the encoder still processes a sequence of the original length but receives no information at masked positions. For Transformer-based architectures, the masking is done by eliminating the masked frames from the sequence. The encoder thus processes only the subsequence of unmasked tokens, significantly reducing computational cost since self-attention operates on a shorter sequence. The masking is done after the positional encodings are added, ensuring that the remaining tokens retain their absolute position information within the original sequence.

\subsection{Training Process}
\label{sec:upstream_training}
\vspace{-1mm}

We segment all data used for training the embedding extractors into Intonation Units (IUs). IU boundaries are detected automatically using the pretrained model from \cite{IUextractor}. This linguistically-motivated segmentation ensures that each input to the autoencoder contains a complete prosodic contour, as opposed to a random fragment, which facilitates learning coherent prosodic representations.

We train the auto-encoders on two English speech corpora.
The first is \textbf{LJSpeech} \cite{LJSPEECH}, a single-speaker corpus of approximately 24 hours of read English speech from a female speaker, consisting of short clips from non-fiction books recorded in a studio. The expressive nature of the reading style provides diverse prosodic patterns. The second is \textbf{VCTK} \cite{VCTK}, a multi-speaker corpus containing approximately 44 hours of clean speech from 110 English speakers with various accents. The multi-speaker setting provides exposure to a wide range of pitch ranges, speaking rates, and prosodic styles. Both corpora are recorded in clean, high-quality conditions, which is critical for reliable extraction of F0 and IU segments.
While larger speech datasets like LibriSpeech are available for pre-training, they are not clean enough for this processing. In future work we plan to explore the use of augmentation or denoising approaches to enhance robustness of the approach. 

For the GRU-based models, we employ two strategies to stabilize training given the variable-length nature of the input sequences. First, we use curriculum learning: the training set is partitioned into three subsets based on sequence duration. Training proceeds in stages, beginning with the shortest sequences and progressively incorporating longer ones. This allows the model to first learn to encode simple, short sequences before tackling longer and more complex ones, reducing early-stage gradient instability caused by long sequences. Second, the autoregressive decoder is trained with linear teacher forcing decay. During training, a Bernoulli trial with parameter $p$ determines whether the ground-truth frame or the decoder's own previous prediction is fed as input. The value of $p$ is linearly annealed to $0$ over $80$ epochs, after which the decoder operates entirely in free-running mode, consuming its own predictions.



\section{Downstream Evaluation Benchmark}
We would like our prosodic embeddings to: 1) contain only prosodic information, and as a consequence, be relatively invariant to speaker and the text content, and 2) contain all the prosodic information relevant for high-level tasks like emotion recognition or intonation classification. It is important to note that complete immunity to speaker and text content is inherently impossible if we want to satisfy the second condition. Speakers differ in their prosodic patterns, so any prosodic embedding will necessarily be affected by the speaker identity. Further, detailed prosodic patterns are inherently tied to text. Discarding all speaker- and text-dependent information would probably imply also loosing relevant information for some tasks, failing condition 2). We then require a benchmark for assessing the degree to which those conditions are satisfied.

The SUPERB-prosody \cite{superb-prosody} benchmark was designed to assess the performance of self-supervised speech models on a variety of prosody-related tasks. Yet, since these tasks are not purely prosodic as they also depend on lexical and speaker cues, the benchmark is not adequate for assessing whether a given embedding representation is purely prosodic. 
In this work, we develop a benchmark for this purpose. 
Below we describe the datasets and evaluation protocols that compose the benchmark.

\subsection{Downstream Databases}
\vspace{-1mm}

We evaluate on three classification tasks. 
The first dataset is the Synthetic Interrogative–Declarative Database (SynthID), a synthetic speech corpus generated for this work using the state-of-the-art text-to-speech system from ElevenLabs.\footnote{\url{https://elevenlabs.io}}  The corpus is designed for classification into interrogative and declarative intonations. Sixteen sentences are constructed so that they are pragmatically natural when pronounced either as a question or as a statement, differing only in the final punctuation mark (“?” vs. “.”). To increase prosodic variability, expressive synthesis parameters provided by the TTS systems were randomly sampled, yielding multiple realizations with controlled variation in speaking style. The dataset contains 608 utterances from 19 synthetic voices, each producing 16 sentences under both interrogative and declarative conditions. The code used to generate the corpus, along with the data, will be publicly released. 

The second dataset used for the experiments is RAVDESS~\cite{RAVDESS}, a speech emotion corpus with 1,440 utterances from 24 speakers (12 female, 12 male). Each speaker produces two fixed statements (``kids are talking by the door'', ``dogs are sitting by the door'') with eight emotions. For each emotion–statement pair, two repetitions are recorded. Additionally, all emotions except neutral are expressed at two intensity levels (normal and strong), whereas neutral is recorded at a single intensity level.

The third dataset is Bestiary \cite{bestiary}, a corpus in which 28 English speakers produced the same target sentences in three pragmatic contexts designed to elicit different intonational contours: contradiction, incomplete response, and incredulity. The classification target is given by the intonational contour actually produced by each speaker, as annotated by four researchers without knowledge of the context, with a very high degree of inter-annotator agreement. The annotated contour categories include falls, verum focus falls, rise-fall-rise (RFR), contradiction contour (CC), yes/no rise (YNR), and other.

In this dataset, not every speaker produced every contour class for every text. This introduces a correlation between text and class. Models may then tend to use text information to determine the class for texts which are mostly uttered with a single contour class. 
To solve this, we discarded samples as follows. First, minority classes were discarded, keeping only YNR, RFR, and CC. Then, speakers were grouped so that each group included a similar amount of samples for each of these three classes. Then, we discarded samples until each text/contour combination was equally represented within each speaker group. This ensures that the text is not informative of the contour class. After balancing, the dataset contains 358 utterances from 19 speakers across 6 speaker groups, 9 target texts, and 3 contour classes (YNR, RFR, and CC).

\subsection{Evaluation protocols}
\vspace{-1mm}

\begin{figure}[t]
    \centering
    \includegraphics[width=0.8\linewidth]{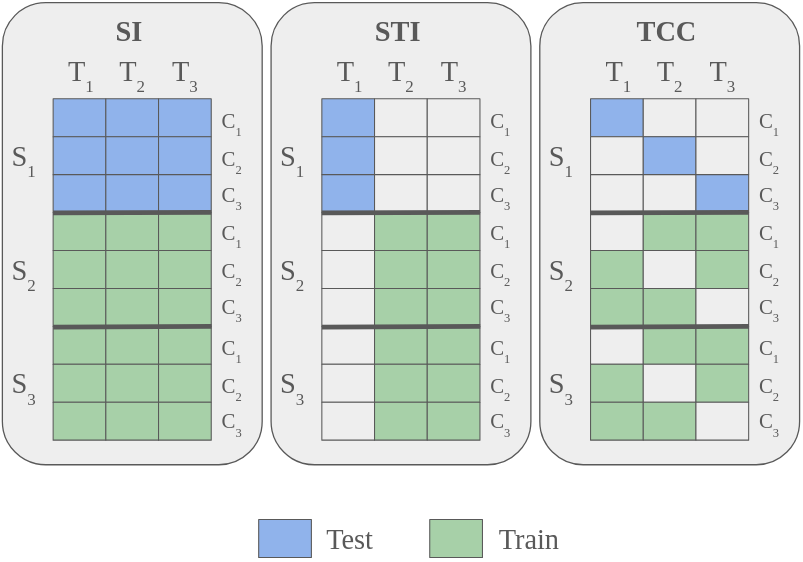}
    \caption{Schematic of the cross-validation approaches used in this work for a toy problem with 3 texts ($T_i$), 3 speakers ($S_i$), and 3 classes ($C_i$), and a single sample per combination, so that each rectangle represents a sample. The green samples are used to train a model which is then applied to the blue samples. This process is repeated by changing the test samples and corresponding training samples until all samples are used for testing exactly once.}
    \label{fig:xval_scheme}
\end{figure}

We define three cross-validation protocols of increasing difficulty, illustrated in Figure~\ref{fig:xval_scheme}:

\textbf{Speaker-Independent (SI)}: The data is split into folds by speaker. This ensures that, during testing, the classifier would not benefit from the use of speaker-specific cues that it may have learned during training.

\textbf{Speaker-Text-Independent (STI)}: In this case, in addition to speaker independence, we assume text independence. That is, we ensure that the specific texts (sentences) in the test set are not seen during training. This assesses the classifier's ability to generalize across both unseen speakers and unseen lexical content. To implement this, the speaker independent folds from above are split into subfolds by text. Then, a model is trained for each subfold using all other subfolds that do not contain the same texts or speakers.

\textbf{Text-label Correlation (TCC)}: This protocol is designed to test robustness to spurious correlations between the text and the target class. To this end, the speaker-independent folds are split into subfolds by text and class. 
Then, when training the model for a subfold corresponding to text T and class C, all other subfolds from different speakers are used, except those with text T and class C. 
In that way, the text/class combinations during testing are not seen by the model during training. 
A classifier that uses features that encode text information should perform poorly under this protocol, since, for a given test sample, it may tend to predict one of the classes seen during training for the test text, which are different from the class of the test sample.  For expedience, in practice, we select several speaker/text/class combinations for testing and train a model discarding that speaker and all the test text/class combinations. 

In all cases, every test sample is scored exactly once, using a model trained on samples that satisfy the restrictions of the protocol. Then,  the scores for all test samples are pooled together to compute the performance metrics. Each experimental configuration is run with 3 different random seeds to account for variability in the training process. The reported performance corresponds to the average across seeds. To estimate statistical uncertainty, we compute 95\% confidence intervals using speaker-conditioned bootstrapping with 100 bootstrap samples.\footnote{Confidence intervals are computed using \url{https://github.com/luferrer/ConfidenceIntervals}} 

\subsection{Downstream Models}

The downstream evaluation is performed using a simple multi-layer perceptron (MLP) taking the fixed-length prosodic embeddings, $z \in \mathbb{R}^d$, as input. The MLP consist of a configurable number of hidden layers with ReLU activations and dropout, followed by a linear output layer. Two configurations were considered: one where all hidden layers have the same width, and one where each successive layer halves the width of the previous one. In both cases, a final output layer projects the last hidden layer dimension to the output dimension for the task.

The optimal architecture and training hyperparameters are selected for each configuration using Bayesian hyperparameter search, optimizing accuracy within the SI protocol using the SynthID dataset. The search space includes the number of hidden layers $\in \{1, 2, 3\}$, hidden layer width $\in \{64, 128, 256, 512, 1024\}$, architectural shape $\in \{\text{straight}, \text{bottleneck}\}$, dropout rate $\in \{0.0, 0.15, 0.3\}$, learning rate $\in \{2\!\times\!10^{-3}, 1\!\times\!10^{-3}, 5\!\times\!10^{-4}, 1\!\times\!10^{-4}\}$, and batch size $\in \{16, 32, 64\}$. The selected architecture is then used for the other two downstream tasks, RAVDESS and Bestiary.

To determine the optimal number of training epochs for the downstream models, we employ nested cross-validation. In the inner cross-validation loop, we always use speaker-independent splits. The optimal epoch is determined within the inner loop by running cross-validation over the training data for that fold, selecting the best epoch, and then retraining a model with that number of epochs using all the training data for that fold.
In practice, we found that setting the maximum number of epochs to 50 was enough to reach stable validation performance.

\section{Baselines}

\label{sec:baselines}

To contextualize the performance of our learned prosodic embeddings, we compare against three categories of baseline representations that differ in their design, information content, and level of abstraction.

\subsection{Hand-crafted prosodic features: eGeMAPS}
The extended Geneva Minimalistic Acoustic Parameter Set (eGeMAPS) \cite{egemaps} is a standardized set of 88 acoustic features extracted using the openSMILE toolkit \cite{eyben2010opensmile}. It comprises statistical functionals (mean, standard deviation, percentiles, slopes, etc.) computed over low-level descriptors including F0, loudness, spectral flux, formant frequencies, and voice quality measures such as jitter, shimmer, and harmonic-to-noise ratio. eGeMAPS was designed by expert consensus to be a compact and interpretable feature set for affective computing and paralinguistic tasks, and remains a widely used baseline in emotion recognition and speaker state analysis \cite{emoref1, emoref2, emoref3}.

For this work, we use the subset of eGeMAPS features that are derived from the pitch and energy signals  and from those signals plus jitter and shimmer (called egemaps-prosody and egemaps-prosody+ in the plots).
For our purposes, eGeMAPS serves as a reference point for fixed, non-learned prosodic summaries. Comparing against eGeMAPS allows us to assess how much can be gained by learning the representation from data.

\subsection{Self-supervised speech representations}
Large self-supervised speech models learn general-purpose representations from raw waveforms and have achieved state-of-the-art results across a wide range of speech tasks. We include two such models as baselines.

\textbf{WavLM} \cite{wavlm} is a Transformer-based model pre-trained on 94k hours of speech using a masked speech prediction objective with a denoising component. It produces frame-level contextualized representations (1024-dimensional at 50 Hz for the Large variant) that capture a rich mixture of acoustic, phonetic, speaker, and prosodic information. WavLM has demonstrated strong performance on the SUPERB benchmark across tasks spanning speech recognition, speaker verification, and emotion recognition. We evaluate two strategies for obtaining utterance-level embeddings from WavLM:
\begin{itemize}
    \item \textbf{WavLM-mean}: The frame-level outputs of all Transformer layers are averaged (across layers), and the resulting per-frame vectors are then mean-pooled over time. This strategy gives equal weight to every layer, combining low-level acoustic information from earlier layers with the more abstract representations in later layers. It follows the spirit of the SUPERB framework \cite{superb}, which showed that different downstream tasks benefit from different layers, suggesting that a cross-layer summary can serve as a versatile general-purpose embedding.
    \item \textbf{WavLM-last}: Only the output of the final Transformer layer is retained and mean-pooled over time. The last layer provides the most contextualized representation, as it aggregates information across the full temporal context and is directly shaped by the pretraining objective. This may be beneficial for modeling high-level prosodic patterns. 
\end{itemize} 

\textbf{Emotion2Vec} \cite{emo2vec} is a speech emotion representation model. It is first pre-trained in a self-supervised fashion using masked acoustic modeling on large-scale emotional speech data: waveforms are encoded via a convolutional frontend and a Transformer encoder, where randomly masked latent frames are reconstructed from contextual representations. After pre-training, the model is fine-tuned for emotion recognition using time-pooled utterance-level embeddings. The resulting representations are specifically tailored to capture affective and paralinguistic information, and have shown strong results on emotion recognition benchmarks. 

These two baselines represent upper-bound references in terms of the information available to the representation: since they operate on the full waveform, they have access to all acoustic cues, including lexical content, speaker timbre, and channel characteristics, in addition to prosody. Strong performance from these models on our downstream tasks is expected, particularly under the Speaker-Independent protocol where text-specific cues can be leveraged. However, under the more challenging Speaker-Text-Independent and Text-Emotion Correlation protocols, the non-prosodic information encoded in these representations may become a liability if the classifier learns to rely on text or speaker cues that do not generalize to unseen texts or speakers. Comparing our prosody-only embeddings against these waveform-based models allows us to assess the value of our proposed approach.

\subsection{Frame-level prosodic embeddings: ProsodyVQ-VAE}
Portes and Hor\'{a}k \cite{prosodyVQVAE2024, prosodyVQVAE2025} proposed a VQ-VAE model that jointly encodes frame-level F0 and energy into a discrete codebook of vector embeddings. Their model uses a convolutional encoder--decoder that downsamples the input by a factor of 16 (from 200\,Hz to 12.5\,Hz) and quantizes the latent representations to the nearest vector in a learned codebook. They conducted an extensive hyperparameter search on LibriTTS and achieved very low reconstruction error (below 1\% FFE for interpolated F0), demonstrating that the codebook faithfully preserves the input signals.

This baseline is the closest to our work in that it also learns representations from prosodic features (F0 and energy) rather than from the full waveform. However, the two approaches differ in fundamental ways. The VQ-VAE produces \emph{frame-level discrete tokens} (one every 80\,ms), whereas our models produce \emph{utterance-level continuous embeddings} that summarize the entire prosodic contour of an intonation unit. The discrete nature of the codebook introduces a limited resolution in feature space that may not be optimal for all downstream tasks.

In the original paper, the model is optimized exclusively for reconstruction fidelity, with no downstream evaluation to assess whether the learned codes capture higher-level prosodic information. Here, we compare it to our proposed approaches on downstream task performance. To this end, we compute a global embedding comparable to ours by extracting the VQ-VAE codebook vectors for each utterance and concatenating the mean and standard deviation over time. 

\section{Results and Discussion}

%

In this section we first show ablation experiments, comparing various configurations on the SynthID dataset where we made all development decisions. Next, we show results for a few selected configurations and the baseline approaches on the two held-out datasets, RAVDESS and Bestiario.

\subsection{Comparison of configurations on the SynthID dataset}

\begin{figure}[t]
    \centering
    \includegraphics[width=0.9\linewidth]
    {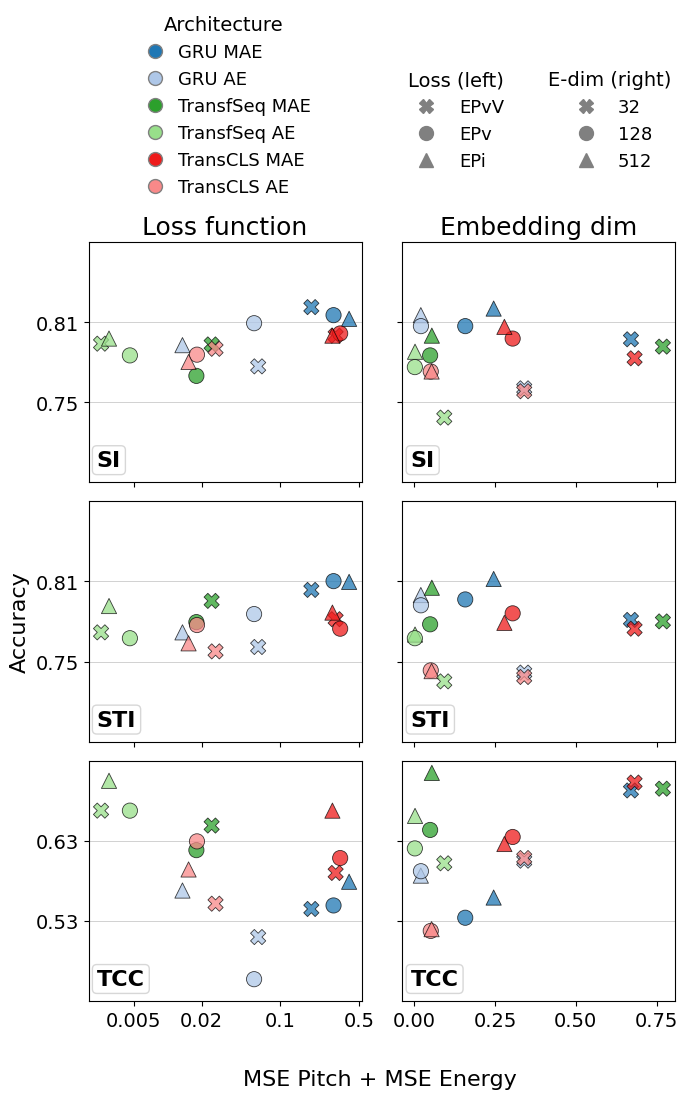}
    \caption{Accuracy and MSE results for various system configurations for the three evaluation protocols on the SynthID dataset. The y-axis range is determined so that it approximately spans the size of three confidence intervals for this dataset. The horizontal lines are separated by the size of the intervals (averaged across systems), visually indicating the statistical significance (or lack thereof) of the difference between the systems. The color of the markers indicates the architecture and pretext task, the shape indicates the loss (left column) or the dimension of the embeddings (right column).}
    \label{fig:task_compar}
\end{figure}

Figure \ref{fig:task_compar} shows the results for the three architectures explored: GRU, Transformer with sequential output (TransfSeq), and with CLS-token (TransfCLS). For each case we show results for two pretex-tasks (standard and masked auto-encoding, AE and MAE, respectively). On the left, we explore those configurations for three loss functions (EPv, EPi, and EPvV), and on the right, for three embedding dimensions (32, 128 and 512). Each row corresponds to one evaluation protocol, showing two metrics: the combined MSE of the pitch and energy signals, and the accuracy value for the SynthID task. 

For the SI and STI protocols, for which there is no spurious correlation between text and class, all configurations lead to very similar performance on the downstream task, despite the fact that reconstruction error is wildly different, spanning a range of 2 orders of magnitude. For TCC, some configurations have significantly different downstream performance from each other. Yet, better (lower) MSE values do not correspond to better (higher) accuracy values. These results suggest that the quality of prosodic representations cannot be assessed using reconstruction metrics. 

Focusing on the plots on the left, where the embedding dimension is fixed to 256, and comparing the trends for the SI and STI cases with those for the TCC case, we can see an almost reversed -- though rather noisy -- pattern: systems that are better for SI and STI are worse for TCC and conversely. The models that pass a single fixed-length embedding to the decoder (GRU and TransfCLS) require the embedding to retain enough detail about the temporal sequence for reconstruction of the input signal. Such detail is useful for the SI and STI conditions where there are no spurious correlations between linguistic content and class. Recurrent reconstruction may be more difficult than parallel one, requiring the embedding to retain more of the details, explaining the superior performance of the GRU for those cases. 

On the other hand, the detailed temporal patterns learned by those models are fragile when spurious correlations between text and class are found in the training data, as for the TCC case. In this scenario, the best approach is one where the decoder has access to a sequence, making the reconstruction task simpler. The embedding in this case is obtained by temporal pooling, which potentially discards some of the fine-grained temporal information that is less robust in this scenario. 

Notably, no clear winner is observed when comparing the three loss types (EPv, EPi, and EPvV) indicating that the use of the loss over the voicing signal or the interpolated unvoiced regions was probably not necessary. Given that all approaches lead to similar results,  we will use only on the EPvV loss for the rest of the experiments in this work.

Focusing now on the plots on the right of Figure \ref{fig:task_compar} where the loss is fixed to EPvV and the embedding dimension is varied, we see that, for most models, the smaller dimension (32) is comparable to the two larger ones in terms of accuracy, while being markedly worse in terms of MSE for the MAE options. In fact, for the TCC case, the smaller dimension for the MAE configurations leads to close-to-optimal results. This extremely compact representation appears to have all relevant prosodic information, while providing robustness to challenging conditions.


\subsection{Comparison with baseline approaches}

\begin{figure*}
    \centering
    \includegraphics[width=0.97\linewidth]{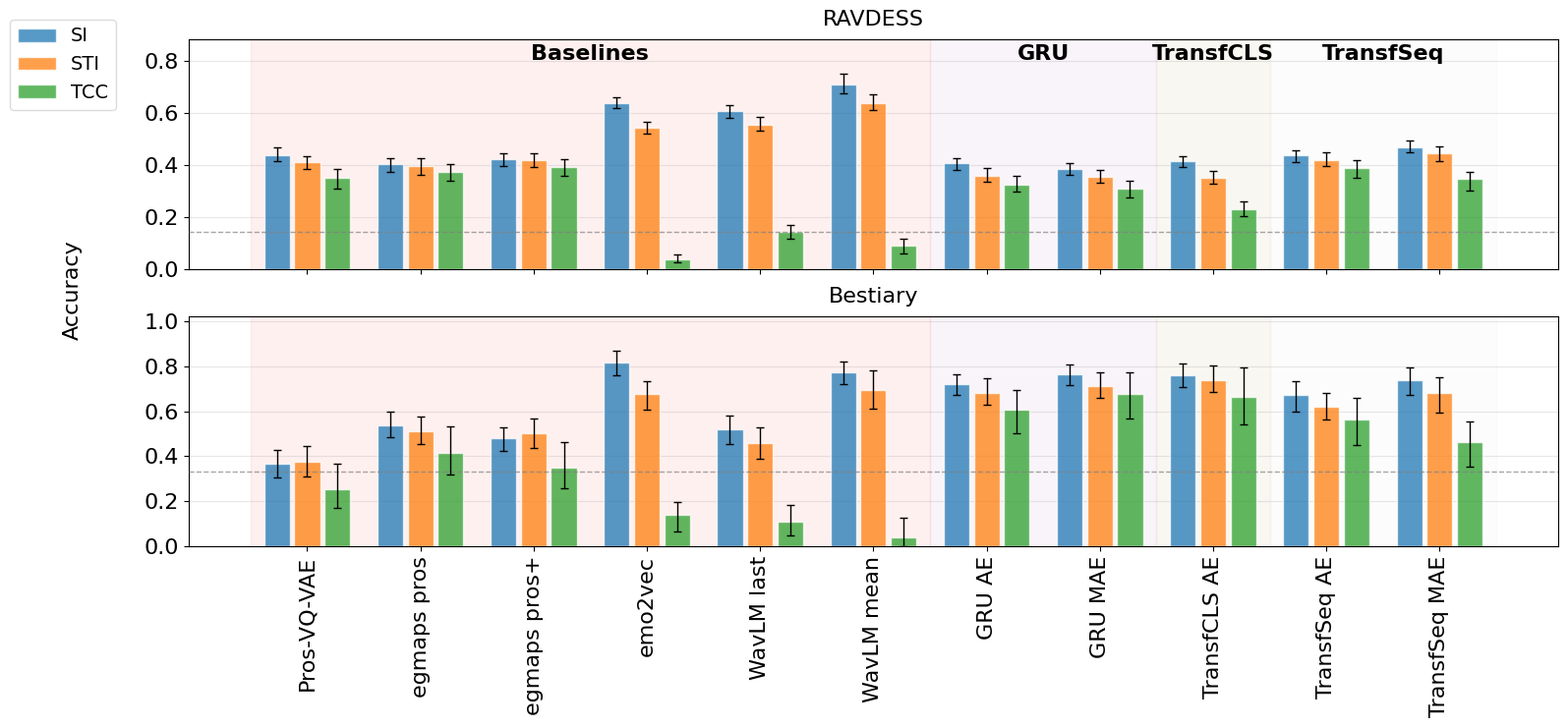}
    \caption{Results on the RAVDESS and Bestiary datasets for various baseline approaches and a subset of configurations for our proposed method corresponding to the EPvV loss and an embedding size of 256. The grey horizontal line indicates the random baseline for accuracy, obtained by predicting always the majority class. The whiskers on top of the bars indicate the 95\% confidence intervals.}
    \label{fig:ravdess_bestiary}
\end{figure*}

Figure \ref{fig:ravdess_bestiary} shows results for the RAVDESS and Bestiario hold-out datasets, where no development decisions were made. We include results for the baseline systems from Section \ref{sec:baselines}. For our proposed approaches, we show results with $d=128$ and EPvV loss for every combination of architecture and pretext task.

On the RAVDESS emotion classification task, waveform-based models that encode rich information about the linguistic content and speaker characteristics (WavLM and emotion2vec) dominate the results for the SI and STI cases. This is expected since emotion recognition benefits not only from prosodic cues but also from voice quality and spectral characteristics that carry speaker-dependent emotional signatures—information that is available to models operating on the full waveform but deliberately excluded from prosodic representations. 

Notably, though, for the TCC case, these features lead to very poor performance, suggesting that the model is at least partly relying on the linguistic content to make predictions, leading to incorrect decisions when the test content and class combination was never seen during training. This also has a small impact in the STI scenario, where whatever linguistic cues were learned by the model generalize poorly to the test scenario where the linguistic content is different, degrading performance with respect to the SI case. This degradation is also observed on some of our embeddings, suggesting that some text-dependent information is present in the prosodic embeddings, a reasonable assumption since prosody and text are unavoidably related.

To quantify the degree to which the different embeddings preserve speaker and linguistic information, we trained logistic linear regression models to predict this information using the embeddings as input on the SynthID dataset. Results show that while for the WavLM mean system the accuracy for speaker and sentence classification are 72\% and 100\%, respectively, for the TransfSeq-AE approach, they are 40\% and 41\%. This shows that our embeddings contain much less speaker and linguistic information, but, as expected, are not completely immune to it. Being invariant to speaker and linguistic changes would imply loosing important prosodic details which we wish to preserve.

The approach that is most directly comparable to ours is ProsodyVQ-VAE, as it also operates on F0 and energy signals. The figure shows that our best configurations consistently outperform this model under the SI and STI protocols. 
For the TCC protocol, 
eGeMAPS and ProsodyVQ-VAE features achieve the best results. The robustness of eGeMAPS is consistent with the fact that these features were designed by expert consensus specifically for affective computing tasks, and their statistical nature (means, percentiles, slopes over functionals) makes them largely agnostic to the specific text realization, as reflected by the small variation in performance across the three evaluation protocols.

Finally, the bottom plot in Figure  \ref{fig:ravdess_bestiary} shows the results for the Bestiary dataset. In this case, the gap between the best baselines and our prosody-only embeddings for the SI and STI cases is notably smaller than for RAVDESS. This is because intonation contour classification is a more purely-prosodic task than emotion recognition, where voice quality and spectral cues also play a significant role. Hence, our prosodic embeddings convey nearly all the information relevant for this task. Crucially, in the presence of spurious class/text correlations (TCC results), the additional information present in the audio-based embeddings results in below-chance results. 

Comparing results with the baseline prosodic representations, egmaps and ProsodyVQ-VAE, our models outperform these systems by a wide margin across all protocols for this task. 
For the TCC protocol, our proposed embeddings achieve the best results on this task, outperforming eGeMAPS.  
The fact that eGeMAPS outperforms our embeddings on the TCC protocol for RAVDESS but not for Bestiary may be explained by the fact that these features consist of statistics over the temporal signals, loosing the detailed information about the temporal patterns. The intonation contour distinctions in Bestiary may not be well represented by these features compared to our learned prosodic representations. 


\section{Conclusions}

We proposed an approach for extraction of global prosodic embeddings from speech signals based on auto-encoder models of pitch and energy signals. We compared GRU- and transformer-based auto-encoders, with  standard or masked reconstruction pretext tasks. Further, we proposed a benchmark for evaluation of prosodic embeddings, consisting of three tasks and three evaluation protocols of increasing difficulty, aiming to assess the degree to which the embeddings 1) preserve prosodic information and 2) are affected by speaker and linguistic content. Purely prosodic embeddings should be relatively robust to speaker and linguistic information -- though complete immunity to linguistic content and speaker identity, while preserving all valuable prosodic information, is impossible due to the inherent correlation between these aspects.   





Our results showed that the proposed method results in embeddings that are relatively robust to mismatches in speaker and linguistic content compared to audio-based embedding extraction methods, while retaining at least as much information for solving prosodic tasks as other purely-prosodic embeddings. 
These embeddings can be used as a drop-in replacement for hand-crafted prosodic features, as input to speech synthesis models, and in combination with lexical or speaker representations for solving task where prosody provides relevant information. They can also be used to study the importance of prosodic information on a given task.
The benchmark, code and models developed for this work are publicly available for research use.


\label{sec:future_work}

\section{Generative AI Use Disclosure}
We used a generative AI tool for light language editing and translation. All experimental design, implementation decisions, analyses, and interpretations were carried out and validated by the authors, who take full responsibility for the work.

\bibliographystyle{IEEEtran}
\bibliography{mybib}

\end{document}